\documentclass[conference]{IEEEtran}
\IEEEoverridecommandlockouts
\usepackage{cite}
\usepackage{mathtools}
\usepackage{amsmath,amssymb,amsfonts}
\usepackage{algorithmic}
\usepackage{graphicx}
\usepackage{balance}
\usepackage{textcomp}
\usepackage[caption=false,font=footnotesize]{subfig}
\usepackage{xcolor}
\usepackage{multirow,multicol}

\DeclarePairedDelimiter\floor{\lfloor}{\rfloor} %f floor operator
\DeclareMathOperator{\EX}{\mathbb{E}}% expected value
\DeclareMathOperator{\tr}{tr} % Trace symbol
\DeclareMathAlphabet\mathbfcal{OMS}{cmsy}{b}{n} % bold mathcal font
\DeclareMathOperator*{\argmin}{arg\,min}
\newcommand{\RNum}[1]{\lowercase\expandafter{\romannumeral #1\relax}}
\usepackage[linesnumbered,ruled,vlined]{algorithm2e} %for algorithms
\makeatletter
\newcommand{\nosemic}{\renewcommand{\@endalgocfline}{\relax}}% Drop semi-colon ;
\newcommand{\dosemic}{\renewcommand{\@endalgocfline}{\algocf@endline}}% Reinstate semi-colon ;
\let\oldnl\nl% Store \nl in \oldnl
\newcommand{\nonl}{\renewcommand{\nl}{\let\nl\oldnl}}% Remove line number for one line
\makeatother
\newcommand\blfootnote[1]{%
	\begingroup
	\renewcommand\thefootnote{}\footnote{#1}%
	\addtocounter{footnote}{-1}%
	\endgroup
}

\newcommand*\squeezespaces[1]{% %% <- #1 is a number between 0 and 1
	\thickmuskip=\scalemuskip{\thickmuskip}{#1}%
	\medmuskip=\scalemuskip{\medmuskip}{#1}%
	\thinmuskip=\scalemuskip{\thinmuskip}{#1}%
	\nulldelimiterspace=#1\nulldelimiterspace
	\scriptspace=#1\scriptspace
}
\newcommand*\scalemuskip[2]{%
	\muexpr #1*\numexpr\dimexpr#2pt\relax\relax/65536\relax
} %% <- based on  https://tex.stackexchange.com/a/198966/156366

\def\BibTeX{{\rm B\kern-.05em{\sc i\kern-.025em b}\kern-.08em
    T\kern-.1667em\lower.7ex\hbox{E}\kern-.125emX}}
\begin{document}
	\bstctlcite{IEEEexample:BSTcontrol}
%\title{Swarm Intelligence-Based Beamforming and Attenuator Optimization in Full-Duplex Massive MIMO for Enhanced Self-Interference Suppression \vspace{-1ex}}
\title{Adaptive Modulus RF Beamforming for Enhanced Self-Interference Suppression in Full-Duplex Massive MIMO Systems \vspace{-1ex}}
%\title{Suppressing Self-Interference in Full-Duplex Massive MIMO Systems: Swarm Intelligence Inspired Adaptive Modulus RF Beamforming \vspace{-1ex}}
%\thanks{This work was supported in part by Huawei Technologies Canada and in part by the Natural Sciences and Engineering Research Council of Canada.}
\author{Mobeen Mahmood, Yuanxing Zhang, Robert Morawski, Tho Le-Ngoc \\ Department of Electrical and Computer Engineering, McGill University, Montreal, QC, Canada \\
	Email: mobeen.mahmood@mail.mcgill.ca, yuanxing.zhang2@mail.mcgill.ca, robert.morawski@mcgill.ca, \\ tho.le-ngoc@mcgill.ca \vspace{-4ex}}

\maketitle

\begin{abstract}
	This study employs a uniform rectangular array (URA) sub-connected hybrid beamforming (SC-HBF) architecture to provide a novel self-interference (SI) suppression scheme in a full-duplex (FD) massive multiple-input multiple-output (mMIMO) system. Our primary objective is to mitigate the strong SI through the design of RF beamforming stages for uplink and downlink transmissions that utilize the \textit{spatial degrees of freedom} provided due to the use of large array structures. We propose a non-constant modulus RF beamforming (NCM-BF-SIS) scheme that incorporates the gain controllers for both transmit (Tx) and receive (Rx) RF beamforming stages and optimizes the uplink and downlink beam directions jointly with gain controller coefficients. To solve this challenging non-convex optimization problem, we propose a swarm intelligence-based algorithmic solution that finds the optimal beam perturbations while also adjusting the Tx/Rx gain controllers to alleviate SI subject to the directivity degradation constraints for the beams. The data-driven analysis based on the measured SI channel in an anechoic chamber shows that the proposed NCM-BF-SIS scheme can suppress SI by around 80 dB in FD mMIMO systems. 
	\vspace{-1em}
\end{abstract}

\blfootnote{This work was supported in part by Huawei Technologies Canada and in part by the Natural Sciences and Engineering Research Council of Canada.}

\IEEEpeerreviewmaketitle
\vspace{-1em}
\section{Introduction}
\vspace{-1ex}
\IEEEPARstart{M}{assive} multiple-input multiple-output (mMIMO) and full-duplex (FD) communications are two pivotal enablers of next-generation wireless networks to meet the ever-increasing demand for the data traffic. Massive MIMO improves system performance via high multiplexing gain by employing a large number of antennas at the base station (BS), which concentrates the radiated energy on the intended user equipment (UE). When compared to half-duplex (HD) communications, FD communications theoretically double the capacity since they allow the simultaneous transmission of uplink and downlink signals in the same frequency and time resources. Thus, FD and mMIMO together can meet the throughput and latency requirements of next-generation wireless networks with limited spectrum resources \cite{FD_mMIMO}. \par 
%
% Massive MIMO uses a large number of antennas at the base station (BS), which focuses the radiated energy on the desired user equipment (UE), and significantly improves the system performance by providing high multiplexing gain \cite{bjornson2017massive}. FD communications enable the simultaneous transmission of uplink and downlink signals in the same frequency and time resources and can theoretically double the capacity when compared to half-duplex (HD) communications \cite{FD_2}. Thus, FD and mMIMO together can fulfill the throughput and latency demands of nex-generation wireless networks with limited spectrum resources \cite{FD_mMIMO}. \par 
The strong self-interference (SI), which is the signal leakage from the transmitter (Tx) output to the receiver (Rx) input and can be 90 to 110 dB greater than the received signal, prevents the realization of the immense potential of FD technology. Since SI can hinder the Rx capability to detect the intended uplink signals, many research efforts have concentrated on either canceling out or at least significantly suppressing SI in FD systems \cite{nwankwo2017survey}. In this regard, different SI suppression/cancellation techniques can be broadly categorized into the following: 1) passive (antenna) isolation; 2) active cancellation (analog and digital); and 3) spatial suppression. In FD communications systems, antenna isolation, analog/digital SI cancellation (SIC), and their combinations have been used frequently to suppress/cancel the strong SI signal below the Rx noise level \cite{SIC_antenna_3,SIC_joint_1,kim2022performance,Gong2023_FD_Antenna}. In the next-generation wireless networks, there is a growing trend toward utilizing an increased number of antennas at BS. For instance, the deployment of 64-256 antenna configurations has been considered by the third generation partnership project (3GPP) \cite{3GPP_mMIMO}. Therefore, it becomes difficult to implement active analog SIC techniques in FD mMIMO systems due to the prohibitively large complexity that occurs from adding more antennas and SI components. For example, the complexity of an analog circuit for an $M \times M$ antenna array becomes $O(M^2)$. \par 
\textit{Spatial suppression} relies on three dimensional (3D) beamforming of mMIMO systems to mitigate the strong SI by exploiting the extra degrees of freedom (DoF) offered by spatial dimensions of multiple Tx and Rx antennas. In particular, the authors in \cite{SIC_mMIMO_digital} presented SoftNull, which completely eliminates the requirement for analog cancelers by suppressing SI via digital beamforming (DBF). However, DBF becomes impractical for FD mMIMO systems with very large array structures due to the unreasonably high cost/complexity, and energy consumption. On the other hand, by lowering the amount of power-hungry RF chains, hybrid beamforming (HBF), which involves the design of both the radio frequency (RF) and baseband (BB) stages, can approach the performance of DBF. Different HBF solutions are presented for enhancing FD transmission \cite{koc2021,FD_Precoding_4,Koc2022_FD_MU,FD_Precoding_5}. In particular, the angle-of-departure (AoD) and angle-of-arrival (AoA) information is used in \cite{koc2021} to propose a hybrid precoding/combining (HPC) technique for a millimeter-wave (mmWave) FD mMIMO system to suppress SI. The authors in \cite{FD_Precoding_4} introduced the HPC for an FD amplify-and-forward (AF) relay using correlated estimation errors to mitigate SI. For the multi-user (MU) FD mMIMO system in \cite{Koc2022_FD_MU}, the non-orthogonal beams are generated to serve multiple users to maximize sum-rate capacity while suppressing the strong SI. Similarly, the authors in \cite{FD_Precoding_5} show that SI can be reduced by around 30 dB through the joint design of the transmit and receive RF beamformer weights, as well as the precoder and combiner matrices. \par  
Most existing studies in FD mMIMO systems consider using fully-connected (FC)-HBF architectures, where each RF chain is connected to all antenna elements. Alternatively, by linking each RF chain to a particular set of antennas, sub-connected (SC)-HBF architecture can lower the quantity of phase shifters (PSs), and therefore the cost/complexity. Moreover, SC-HBF architectures are better suited for practical implementation in sub-Terahertz (THz) and millimeter-wave (mmWave) communications because they can significantly reduce power consumption at the expense of some performance degradation, but they can also offer a better spectral-energy efficiency tradeoff \cite{alexandropoulos2022full}. A uniform linear array (ULA) SC-HBF architecture is suggested in \cite{mahmood2023subarray} as a means of reducing SI in FD mMIMO systems. A uniform rectangular array (URA) SC-HBF, as opposed to the ULA SC-HBF, can, however, offer better performance in FD mMIMO systems due to the following factors: 1) simultaneous illumination of both uplink and downlink UE azimuth and elevation angles; 2) improved spatial resolution and beamforming flexibility; 3) interference mitigation; and 4) scalability\footnote{Large number of antennas can be accomodated in small space in URA when compared to ULA, which significantly enhances the spatial efficiency.}\cite{mobeen2020,mahmood2021energy,mobeen_3D}.  \par 
In this paper, we propose a novel SI suppression scheme in FD mMIMO systems using a URA SC-HBF architecture. Particularly, the proposed non-constant modulus RF beamforming-based SI suppression (NCM-BF-SIS) scheme optimizes the uplink and downlink beam directions jointly with Tx/Rx variable gain controllers while adhering to the directivity degradation constraints. Our objectives here are twofold: first, to show that the beam perturbation combined with tuned Tx/Rx gain controllers in RF beamformers design can significantly improve SI suppression and can bring the SI level close to the noise floor; and second, to show that the use of URA SC-HBF architecture can provide better performance than ULA SC-HBF. To reduce the high computational complexity during the search for optimal perturbations, we propose a swarm intelligence-based algorithmic solution to find the optimal perturbations, and adjusting the Tx/Rx gain controllers to minimize SI while satisfying the directivity degradation constraints for the uplink and downlink beams. We consider a data-driven analysis of the measured SI channel in an anechoic chamber for over a bandwidth of 20 MHz (centered at 3.5 GHz). The results show that the proposed scheme can suppress SI upto 80 dB in real-time implementations. Morevoer, the use of URA SC-HBF can provide an average SI suppression gain of around 9 dB when compared to ULA SC-HBF.
%the constant modulus RF beamforming-based SI suppression (CM-BF-SIS) scheme optimizes the uplink and downlink beam directions (i.e., azimuth and elevation angles) whereas, 
%----------------------------Figure 1 ---------------------
\begin{figure}[!t]
	\centering
	\captionsetup{justification=centering}
	\includegraphics[height= 5.5cm, width=1\columnwidth]{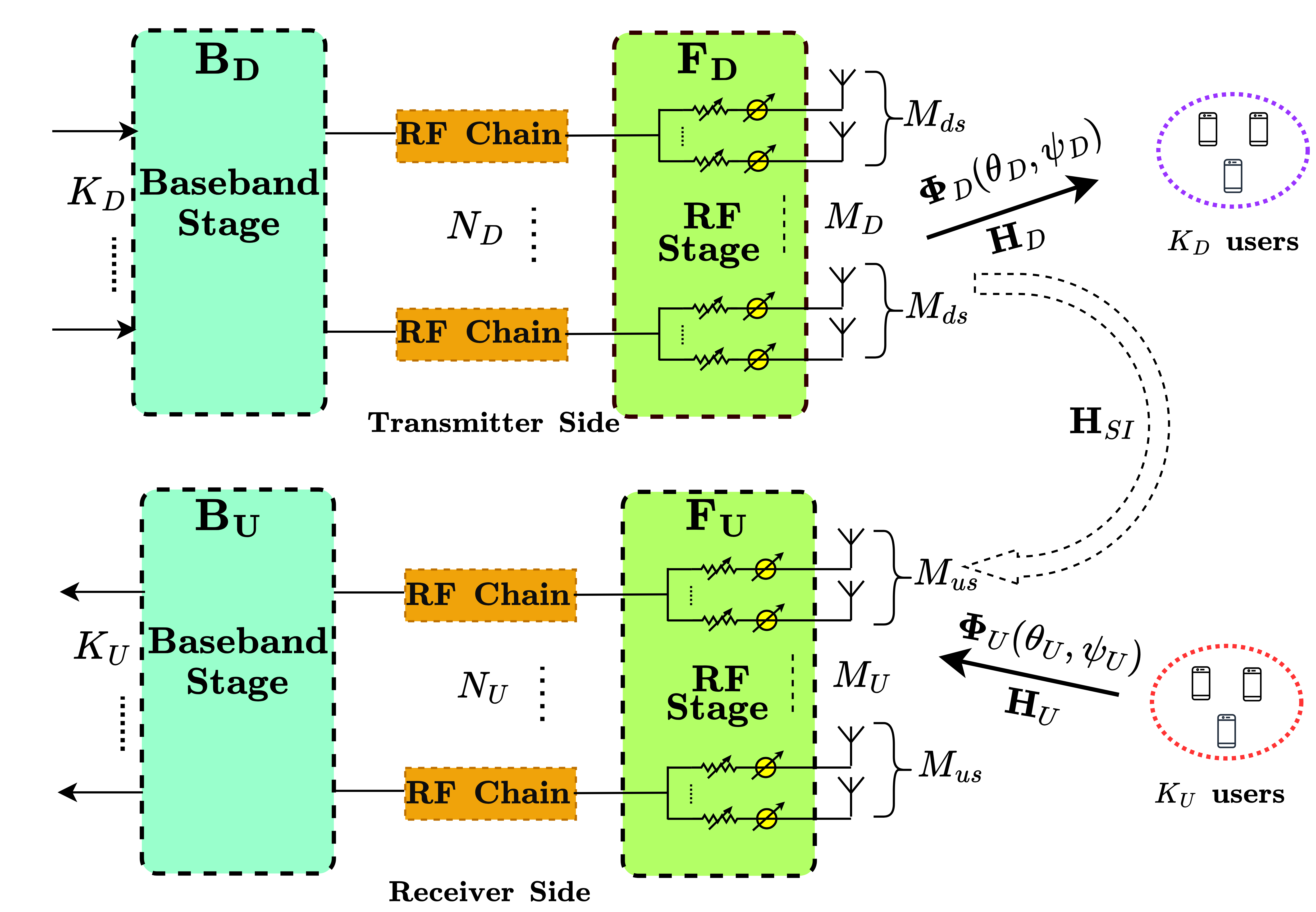} 
		\vspace{-3ex}
	\caption{System model of FD mMIMO HBF communications system.}
	\label{fig:fig1}
	\vspace{-5ex}
\end{figure} 
%--------------------------------------------------------
\section{System Model \& Measured SI Channel}
\subsection{System Model}
We consider a single-cell FD mMIMO system for joint uplink and downlink transmission as shown in Fig. \ref{fig:fig1}. The BS operates in FD mode to simultaneously serve $K_D$ ($K_U$) downlink (uplink) single-antenna UEs over the same frequency band, while the UEs operate in HD mode due to the hardware/software constraints on UEs (e.g., low power consumption, limited signal processing and active/passive SI suppression capability). The BS is equipped with Tx and Rx URAs. Specifically, the Tx (Rx) URA has $M_D=M_D^{(x)} \times M_D^{(y)} (M_U=M_U^{(x)} \times M_U^{(y)})$ antennas, where $M_D^{(x)} (M_U^{(x)})$ and $M_D^{(y)}(M_U^{(y)})$ denote the numbers of Tx (Rx) antennas along $x$-axis and $y$-axis, respectively.  \par 
For the proposed FD mMIMO system, we consider the downlink signal processed through BB stage $\mathbf{B}_D \in \mathbb{C}^{N_D \times K_D}$ and RF beamformer $\mathbf{F}_D \in \mathbb{C}^{M_D \times N_D}$, where $N_D$ is the number of RF chains such that $K_D \leq N_D \ll M_D$. Similarly, the received uplink signal at BS is processed through RF combiner $\mathbf{F}_U \in \mathbb{C}^{N_U \times M_U}$ and BB combiner $\mathbf{B}_U \in \mathbb{C}^{K_U \times N_U}$ by utilizing $K_U \leq N_U \ll M_U$ RF chains. Here, the RF beamforming stages (i.e., $\mathbf{F}_U$ and $\mathbf{F}_D$) are built using low-cost PSs and variable gain controllers. The downlink channel matrix is denoted as $\mathbf{H}_D \in \mathbb{C}^{K_D \times M_D}$ with $\mathbf{h}_{D, k} \in \mathbb{C}^{M_D}$ as the $k^{th}$ DL UE channel vector. Similarly, $\mathbf{H}_U \in \mathbb{C}^{M_U \times K_{U}}$ is the uplink channel matrix with $\mathbf{h}_{U, k} \in \mathbb{C}^{M_U}$ as the $k^{th}$ uplink UE channel vector. Due to the FD transmission, the SI channel matrix $\mathbf{H}_{S I} \in \mathbb{C}^{M_U \times M_D}$ is present between Tx and Rx antennas at the BS. Then, the transmitted signal vector at the BS is defined as $\mathbf{s}_D=\mathbf{F}_D \mathbf{B}_D \mathbf{d}_D \in \mathbb{C}^{M_D}$, where $\mathbf{d}_D=\left[d_{D, 1}, \cdots, d_{D, K_D}\right]^T \in \mathbb{C}^{K_D}$ is the downlink data signal vector such that $\EX\{\mathbf{d}_D \mathbf{d}_D^H\}=\mathbf{I}_{K_D}$. The transmitted signal vector satisfies the maximum downlink transmit power constraint, which is $\EX\{||\mathbf{s}_D||^2\}=\tr(\mathbf{F}_D \mathbf{B}_D \mathbf{B}_D^H \mathbf{F}_D^H) \leq P_D$, where $P_D$ is the total downlink transmit power. Then, the received downlink signal vector is given as follows: 
\begin{equation}
	\mathbf{r}_D=\underbrace{\mathbf{H}_D \mathbf{F}_D \mathbf{B}_D \mathbf{d}_D}_{\text{Desired Signal}}+ \underbrace{\mathbf{H}_{U}\mathbf{d}_U}_{\text{IUI by uplink UE}}  + \underbrace{\mathbf{w}_D}_{\text{Noise}},
\end{equation}
where $\mathbf{H}_U \in \mathbb{C}^{K_D \times K_U}$ is the inter-user interference (IUI) between the downlink and uplink UE, and $\mathbf{w}_D=\left[w_{D, 1}, \cdots, w_{D, K_D}\right]^T$ $\sim$ $\mathcal{C N}\left(0, \sigma_W^2 \mathbf{I}_{K_D}\right)$ is the complex circularly symmetric Gaussian noise vector. Here, we define $P_U$ as the transmit power of each uplink UE. Similar to the downlink data signal vector, the uplink received signal at BS can be written as:
\begin{equation}
	\tilde{\mathbf{r}}_U\hspace{-0.5ex}=\hspace{-0.5ex}\underbrace{\hspace{-0.35ex}\mathbf{B}_U\hspace{-0.35ex} \mathbf{F}_U\hspace{-0.35ex} \mathbf{H}_U\hspace{-0.35ex} \mathbf{d}_U\hspace{-0.35ex}}_{\text{Desired Signal}} + \underbrace{\hspace{-0.35ex}\mathbf{B}_{U}\hspace{-0.35ex}\mathbf{F}_{U}\hspace{-0.35ex}\mathbf{H}_{SI}\hspace{-0.35ex}\mathbf{F}_{D}\mathbf{B}_{D}\hspace{-0.15ex}\mathbf{d}_{D}\hspace{-0.15ex}}_{\text{SI}}\hspace{-0.5ex} + \hspace{-3ex}\underbrace{\tilde{\mathbf{w}_U}}_{\text{Modified Noise}},
\end{equation}
where $\mathbf{d}_U=\left[d_{U, 1}, \cdots, d_{U, K_U}\right]^T \in \mathbb{C}^{K_U}$ is the uplink data signal vector such that $\mathrm{E}\left\{\mathbf{d}_U \mathbf{d}_U^H\right\}=\mathbf{I}_{K_U}$ and $\tilde{\mathbf{w}_U}=\mathbf{B}_U \mathbf{F}_U\mathbf{w}_U$, where $\mathbf{w}_U = \left[w_{u, 1}, \cdots, w_{U, K_U}\right]^T \sim$ $\mathcal{C N}(0, \sigma_W^2 \mathbf{I}_{K_U})$ is the complex circularly symmetric Gaussian noise vector. The desirable downlink (uplink) beam direction has azimuth and elevation angles $\psi_D (\psi_U)$ and $\theta_D (\theta_U)$, respectively. We define the phase-response vector as:
\begin{equation} \label{eq:phase_response}
	\begin{split}
		&\mathbf{\Phi}(\theta, \psi, M^{(x)}, M^{(y)}) = \\
		& \left[1, e^{j2\pi d \sin(\theta) \cos(\psi)}, \ldots, e^{j2\pi d (M^{(x)}-1) \sin(\theta) \cos(\psi)}\right]^T \\
		& \otimes \left[1, e^{j2\pi d \sin(\theta) \sin(\psi)}, \ldots, e^{j2\pi (M^{(y)}-1) \sin(\theta) \sin(\psi)}\right]^T,
	\end{split}
\end{equation}
where $\theta$ ($\psi$) represents the elevation (azimuth) angle, $M^{(x)}$, ($M^{(y)}$) denote the number of antennas along the $x$ ($y$)-axis, $d$ is the antenna spacing, and $\otimes$ is the Kronecker product.
\subsection{SI Channel Measurement Setup}
The measurement setup was made in an anechoic chamber (i.e., without external surrounding reflections) and consists of 64 Tx and 64 Rx antenna elements\footnote{Due to limited space, the details of SI channel measurement setup will be discussed in the extended version of this paper.}, which are arranged in the form of URA ($8$$\times$$8$ configuration). The SI channel is mainly due to \textit{internal} coupling between Tx and Rx antenna elements (i.e., consisting of only line-of-sight (LoS) path components). Then, the SI channel is measured for 1601 sampling points between frequency range from 3 GHz to 4 GHz (i.e., over a bandwidth of 1 GHz) such that the complete SI channel matrix $\mathbf{H}_{SI}$ has dimensions of 64 $\times$ 64 $\times$ 1601. Based on the number of antenna elements in both Tx and Rx sub-arrays, the corresponding SI channels can be represented as $\mathbf{H}_{SI}^{(sub)}$ $\in$ $\mathbb{C}^{M_{us} \times M_{ds} \times 1601}$, where, $M_{ds} (M_{us}) = \frac{M_D}{L_D} \big(\frac{M_U}{L_U}\big)$ represents the number of Tx (Rx) antenna elements in $l_d^{th} (l_u^{th})$ sub-array for $l_d = 1, \cdots, L_D (l_u = 1,\cdots, L_U)$. As per 3GPP specification, the uplink and downlink channel bandwidth can vary from 5 MHz to 100 MHz \cite{3GPP_BW}, then the corresponding SI channel for the given BW can be written as: $\mathbf{H}_{SI,B}^{(sub)}$ = $\mathbf{H}_{SI}^{(sub)}(:,:,n)$ $\in$ $\mathbb{C}^{i \times i \times n}$, where $i = \{M_{ds},M_{us}\}$, $B$ is the given bandwidth, and $n = 1,2,\dots, N$ is the sample frequency point selected from a total of $N$ frequency points for a given bandwidth. For instance, the 20 MHz band constitutes $n =1,2,\dots,33$ points for the frequency range from 3.49 GHz to 3.51 GHz. Similarly, for the band of 100 MHz, $n =1,2,\dots,161$ points in the frequency range from 3.45 GHz to 3.55 GHz are considered. 
\vspace{-1ex}
\section{URA SC-HBF Architecture and Problem Formulation for SI Suppression}
\subsection{URA SC-HBF Architecture}
We consider a URA SC-HBF architecture, where the Tx (Rx) URA is divided into $L_D (L_U)$ different sub-arrays in the form of URA. Hence, compared to $M_D \times N_D$ PSs for Tx ($M_U \times N_U$ PSs for Rx), only $M_{ds} \times N_D (M_{us} \times N_U)$ PSs are required as each Tx (Rx) RF chain is connected to $\floor*{\frac{M_D}{L_D}} \big(\floor*{\frac{M_U}{L_U}}\big)$ Tx (Rx) antennas, where $\floor*{.}$ represents the largest integer less than or equal to a real number. Then, the downlink and uplink phase response vectors are given as:
\begin{equation} \label{eq:phase_response_downlink}
	\begin{split}
		& \mathbf{\Phi}_D(\theta_D, \psi_D, M_{\text{ds}}^{(x)}, M_{\text{ds}}^{(y)}) = \mathbf{\Phi}(\theta_D, \psi_D, M_{\text{ds}}^{(x)}, M_{\text{ds}}^{(y)}), \\
		& \mathbf{\Phi}_U(\theta_U, \psi_U, M_{\text{us}}^{(x)}, M_{\text{us}}^{(y)}) = \mathbf{\Phi}^H(\theta_U, \psi_U, M_{\text{us}}^{(x)}, M_{\text{us}}^{(y)}),
	\end{split}
\end{equation}
where $\mathbf{\Phi}_D$($\mathbf{\Phi}_U$) is the Tx(Rx) phase response vector as given in (\ref{eq:phase_response}). By controlling the phase of the signals transmitted or received by the array elements, we can effectively steer the beam in the desired uplink/downlink UE direction, and maximize the radiated energy while minimizing interference from other directions. We consider the MU scenario using a single uplink and a single downlink UE (i.e., $K_D = N_D = 1$, $K_U = N_U = 1$)\footnote{This paper presents a simple scenario with a single uplink and downlink UE to investigate the SI suppression in FD mMIMO systems. However, the proposed scheme can be applied to multiple uplink and multiple downlink UEs, which is left as our future work.}. Moreover, the Tx (Rx) RF beamformer constitutes gain controllers, which scales the amplitudes of the signals transmitted (received) from the antennas as shown in Fig. \ref{fig:fig1}. Then, the RF beamformer $\mathbf{f}_D$ for the single downlink UE can be written as:
\begin{equation} \label{eq:F_D_with_atten}
\mathbf{f}_D \hspace{-0.65ex}= \hspace{-0.65ex}\frac{1}{\sqrt{M_{ds}^{(x)}\hspace{-0.5ex} M_{ds}^{(y)}}}\hspace{-0.25ex} \mathbf{\Phi\hspace{-0.5ex}}_D\hspace{-0.35ex}\big(\hspace{-0.35ex}\theta_D,\hspace{-0.35ex} \psi_D, \hspace{-0.35ex}M_{ds}^{(x)},\hspace{-0.35ex} M_{ds}^{(y)}\hspace{-0.25ex}\big) \hspace{-0.4ex}\odot \hspace{-0.2ex}\mathbf{G\hspace{-0.25ex}}_D \hspace{-0.65ex} \in \hspace{-0.65ex}\mathbb{C}^{M_{ds} \times 1},
\end{equation}
where \mbox{$\squeezespaces{0.1} \mathbf{G}_D \hspace{-0.25ex} = [g_1^{(x)}, \ldots, g_{M_{ds}}^{(x)}, g_1^{(y)}, \ldots, g_{M_{ds}}^{(y)}]^T $} is the vector containing the gain values for Tx URA with $g_i^{(x)}$ ($g_j^{(y)}$) representing the gain value of $i^{th} (j^{th})$ antenna for $i = 1, \cdots, M_{ds}^{(x)} \big(j = 1, \cdots, M_{ds}^{(y)}\big)$. Similarly, the uplink RF beamformer $\mathbf{f}_U$ can be expressed as follows:
\begin{equation} \label{eq:F_U_with_atten}
	\mathbf{f}_U \hspace{-0.65ex}= \hspace{-0.65ex}\frac{1}{\sqrt{M_{us}^{(x)}\hspace{-0.5ex} M_{us}^{(y)}}}\hspace{-0.25ex} \mathbf{\Phi\hspace{-0.5ex}}_U\hspace{-0.35ex}\big(\hspace{-0.35ex}\theta_U,\hspace{-0.35ex} \psi_U, \hspace{-0.35ex}M_{us}^{(x)},\hspace{-0.35ex} M_{us}^{(y)}\hspace{-0.25ex}\big) \hspace{-0.4ex}\odot \hspace{-0.2ex}\mathbf{G\hspace{-0.25ex}}_U \hspace{-0.65ex} \in \hspace{-0.65ex}\mathbb{C}^{1 \times M_{us}},
\end{equation}
where \mbox{$\squeezespaces{0.1} \mathbf{G}_U \hspace{-0.25ex} = [g_1^{(x)}, \ldots, g_{M_{us}}^{(x)}, g_1^{(y)}, \ldots, g_{M_{us}}^{(y)}] $} is the vector containing the gain values for Rx URA with $g_p^{(x)}$ ($g_q^{(y)}$) is the gain value of $p^{th} (q^{th})$ antenna for $p = 1, \cdots, M_{us}^{(x)} \big(q = 1, \cdots, M_{us}^{(y)}\big)$.
\vspace{-2em}
\subsection{Problem Formulation}
\vspace{-1ex}
Based on the downlink and uplink RF beamforming stages, we can write the total achieved SI as:
\begin{equation}
	\textrm{A}_{\textrm{SI}} = -10 \log_{10}\Big(\frac{1}{N}\sum_n \big|\mathbf{f}_U^T\mathbf{H}_{SI}^{(sub)}(:,:,n)\mathbf{f}_D \big|^2\Big). \label{eq:SIC}
\end{equation}
By steering the uplink and downlink beams to the desirable directions (i.e. $\hat{\theta}_U = \theta_U$, $\hat{\theta}_D = \theta_D$, $\hat{\psi}_U = \psi_U$, $\hat{\psi}_D = \psi_D$), the Tx and Rx directivities reach their maxima as follows:
\begin{equation} \label{eq:directivity}
	\begin{split}
	|\mathbf{\Phi}_D^T(\theta_D, \psi_D, M_{ds}^{(x)}, M_{ds}^{(y)})\mathbf{f}_D(\theta_D,\psi_D,M_{ds}^{(x)}, M_{ds}^{(y)})|^2 = M_{ds},  \\ 
	|\mathbf{f}_U^T(\theta_U, \psi_U, M_{us}^{(x)}, M_{us}^{(y)})\mathbf{\Phi}_U(\theta_U, \psi_U, M_{us}^{(x)}, M_{us}^{(y)})|^2 =  M_{us}. 
	\end{split}
\end{equation}
For a FD mMIMO system consisting of RF beamformers $\mathbf{f}_D$ and $\mathbf{f}_U$ with variable gain controllers, and using URA SC-HBF architecture at Tx and Rx of BS, the SI can be minimized by finding the uplink and downlink perturbation angles $\{(\hat{\theta}_U, \hat{\theta}_D), (\hat{\psi}_U, \hat{\psi}_D)\}$ jointly with Tx/Rx variable gain controller coefficient optimization. Then, we can formulate the joint optimization problem for SI suppression as: \vspace{-1ex}
\begin{gather} \label{eq:optimization_problem}
	\begin{align} 
		&\min_{ \left\{ \hspace{-0.5ex}\substack{\hat{\theta}_D, \hat{\theta}_U, \hat{\psi}_D, \hat{\psi}_U \\ g_1, \cdots, g_{M_{ds}} \\ g_1, \cdots, g_{M_{us}}}\hspace{-0.35ex}\right\}} \hspace{-3ex}\quad \frac{1}{N}\hspace{-0.75ex}\sum_n \hspace{-0.5ex} \big|\hspace{-0.25ex}\mathbf{f}_{\hspace{-0.2ex}U}^T(\hspace{-0.35ex}\hat{\psi}_{\hspace{-0.2ex}U}, \hspace{-0.35ex}\hat{\theta}_{\hspace{-0.2ex}U}\hspace{-0.25ex}, \hspace{-0.35ex}\mathbf{G}_{\hspace{-0.2ex}U}\hspace{-0.25ex})\hspace{-0.35ex}\mathbf{H}_{\hspace{-0.2ex}SI}^{\hspace{-0.35ex}(sub)\hspace{-0.35ex}}(:,:,n)\hspace{-0.35ex}\mathbf{f}_{\hspace{-0.2ex}D}(\hspace{-0.35ex}\hat{\psi}_{\hspace{-0.2ex}D}\hspace{-0.25ex}, \hspace{-0.35ex}\hat{\theta}_{\hspace{-0.2ex}D}\hspace{-0.25ex}, \hspace{-0.35ex}\mathbf{G}_{\hspace{-0.2ex}D}\hspace{-0.35ex}) \hspace{-0.35ex}\big|^2 \notag \\
		&\hspace{2ex}\textrm{s.t.} \hspace{1ex} C_1\hspace{-0.35ex}: \hspace{-0.35ex}M_{ds}\hspace{-0.65ex} -\hspace{-0.65ex}|\hspace{-0.15ex}\mathbf{\Phi}_{\hspace{-0.2ex}D}^T\hspace{-0.35ex}(\theta_{\hspace{-0.2ex}D}, \psi_{\hspace{-0.2ex}D}, M_{ds})\mathbf{f}_{\hspace{-0.2ex}D}(\hspace{-0.35ex}\hat{\psi}_{\hspace{-0.2ex}D}\hspace{-0.25ex}, \hspace{-0.35ex}\hat{\theta}_{\hspace{-0.2ex}D}\hspace{-0.25ex}, \hspace{-0.35ex}\mathbf{G}_{\hspace{-0.2ex}D}\hspace{-0.35ex},M_{ds})\hspace{-0.35ex}|^2 \leq \epsilon,  \notag \\
		&\hspace{5.5ex} C_2\hspace{-0.35ex}: \hspace{-0.35ex}M_{us}\hspace{-0.65ex} -\hspace{-0.65ex}|\hspace{-0.15ex}\mathbf{f}_{\hspace{-0.2ex}U}^T(\hspace{-0.35ex}\hat{\psi}_{\hspace{-0.2ex}U}\hspace{-0.25ex}, \hspace{-0.35ex}\hat{\theta}_{\hspace{-0.2ex}U}\hspace{-0.25ex}, \hspace{-0.35ex}\mathbf{G}_{\hspace{-0.2ex}U}\hspace{-0.35ex},M_{us})\mathbf{\Phi}_{\hspace{-0.2ex}U}\hspace{-0.35ex}(\theta_{\hspace{-0.2ex}U}, \psi_{\hspace{-0.2ex}U}, M_{us})\hspace{-0.35ex}|^2 \leq \epsilon,  \notag \\
%		&\hspace{2em} \hspace{4.5ex} C_2: \hspace{1ex}  M_U-|\mathbf{f}_U^T(\hspace{-0.35ex}\hat{\psi}_U, \hspace{-0.35ex}\hat{\theta}_U\hspace{-0.25ex}, \hspace{-0.35ex}\mathbf{G}_U\hspace{-0.25ex})\mathbf{\Phi}_U(\theta_U, \psi_U)|^2 \leq \epsilon, \notag \\
		&\hspace{5.5ex} C_3: \hspace{1ex}  \hat{\psi}_D, \hat{\psi}_U, \hat{\theta}_D, \hat{\theta}_U \in [0, 2\pi],  \notag \\
		&\hspace{5.5ex} C_4: \hspace{1ex}  g_1, \cdots, g_{M_{ds}} \in [0, 1], \notag \\
		&\hspace{5.5ex} C_5: \hspace{1ex}  g_1, \cdots, g_{M_{us}} \in [0, 1], 
	\end{align} 
\end{gather} 
where $C_1$ and $C_2$ refer to the directivity degradation constraints in downlink and uplink UE directions, respectively, i.e., to limit the directivity degradation from the main beam directions $\{(\theta_U, \theta_D), (\psi_U, \psi_D)\}$ to a small value $\epsilon$. The constraint $C_3$ limits the perturbed angles range between 0 and $2\pi$, whereas the constraints $C_4$ and $C_5$ confine the Tx and Rx variable gains within the continuous range from 0 to 1. The optimization problem defined in (\ref{eq:optimization_problem}) is non-convex and intractable due to the non-linearity constraints.
\vspace{-1ex}
\section{Proposed Non-constant Modulus RF Beamforming-Based SI Suppression Scheme}
\vspace{-1ex}
We propose a particle swarm optimization (PSO)-based SI minimization scheme, which optimizes the Tx and Rx variable gain controllers coefficients jointly with the optimal uplink and downlink beam directions $\hat{\theta}_D, \hat{\theta}_U$ while satisfying the constraints of directivity degradation ($C_1$ and $C_2$), uplink and downlink perturbed beam search space ($C_3$), and Tx/Rx gain controller values ($C_4$ and $C_5$). The algorithm starts with a swarm of total $P$ particles, each with its own position, velocity, and fitness value, which are randomly placed in optimization search space of perturbation coefficients. During a total of $T$ iterations, the particle $p$ communicates with others, and move for the exploration of the optimization space to find the optimal solution. We define the perturbation vector $\mathbf{X}_p^{(t)}$ as: \vspace{-0.5ex}
\begin{equation}
	\mathbf{X}_p^{(t)} = [\hat{\theta}_D^p, \hat{\theta}_U^p, \hat{\psi}_D^p, \hat{\psi}_U^p, g_1^p, \cdots, g_{M_{ds}}^p, g_1^p, \cdots, g_{M_{us}}^p], \label{eq:PSO} \vspace{-0.5ex}
\end{equation}
where $p = 1,\dots, P$ and $t = 0,1, \dots, T$. For each $p^{th}$ particle, by substituting (\ref{eq:PSO}) in (\ref{eq:F_D_with_atten}) and (\ref{eq:F_U_with_atten}), the downlink and uplink RF beamformers $\mathbf{f}_D(\mathbf{X}_p^{(t)})$ and $\mathbf{f}_U(\mathbf{X}_p^{(t)})$ can be obtained as function of perturbation angles ($\hat{\theta}_D^p$, $\hat{\psi}_D^p$) and ($\hat{\theta}_U^p$, $\hat{\psi}_U^p$), and gain controller coefficients ($g_1^p, \cdots, g_{M_{ds}}^p, g_1^p, \cdots, g_{M_{us}}^p$) respectively. 
\begin{algorithm}[t!]\label{algo:3}
	\nonl \textbf{Input}: \hspace{-0.5ex} $P, T$, $\mathbf{H}_{SI}^{\hspace{-0.2ex}(sub)}$, $(\theta_D, \psi_D)$, $(\theta_U, \psi_U)$. \\
	\nonl \textbf{Output}: \hspace{-0.5ex}$ \hat{\theta}_D, \hat{\theta}_U, \hat{\psi}_D, \hat{\psi}_U, g_1, \cdots, g_{M_{ds}}, g_1, \cdots, g_{M_{us}}$. \\
	\SetAlgoLined 
	\For{$t = 0:T$}{
		\For{$p = 1:P$}{
			\uIf{t = 0}{
				Initialize the velocity as $\mathbf{w}_{p}^{(0)} = \bf{0}$. \\
				Initialize $\mathbf{X}_p^{(t)}$ uniformly distributed in $[\mathbf{X}_{\text{Low}}, \mathbf{X}_{\text{Upp}}]$. \\}
			\Else {
				Update the velocity $\mathbf{w}_{p}^{(t)}$ via (\ref{eq:velocity}).\\
				Update the vector $\mathbf{X}_{p}^{(t)}$ via (\ref{eq:position}).\\
			}
			Find the personal best $\mathbf{X}_{p,\mathrm{best},n}^{(t)}$ via (\ref{eq:personal_best_PSOLPA}).
		}
		Find the global best $\mathbf{X}_{\text{best}}^{(t)}$ as in (\ref{eq:global_best_PSOLPA}).
	}
	\caption{Proposed NCM-BF-SIS Algorithm} 
\end{algorithm}
\setlength{\textfloatsep}{-4pt} \hspace{-1ex}
By using (\ref{eq:SIC}), we can write the total achieved SI suppression as follows:
\begin{equation}
\hspace{-1ex}\text{A}_{\hspace{-0.15ex}\textrm{SI}}\hspace{-0.2ex}(\mathbf{X}_{\hspace{-0.2ex}p}^{(t)}\hspace{-0.5ex}) \hspace{-0.75ex}=\hspace{-0.75ex} -\hspace{-0.2ex}10\hspace{-0.3ex}\log_{\hspace{-0.25ex}10\hspace{-0.25ex}}\hspace{-0.6ex}\Big(\hspace{-0.5ex}\frac{1}{N}\hspace{-0.95ex}\sum_n\hspace{-0.4ex} \big|\hspace{-0.3ex}\mathbf{f}_{\hspace{-0.2ex}U}^T\hspace{-0.25ex}(\hspace{-0.25ex}\mathbf{X}_{\hspace{-0.2ex}p}^{(t)}\hspace{-0.35ex})\mathbf{H}_{\hspace{-0.35ex}SI}^{\hspace{-0.4ex}(sub)}\hspace{-0.35ex}(:\hspace{-0.25ex},\hspace{-0.25ex}:\hspace{-0.15ex},\hspace{-0.25ex}n\hspace{-0.15ex})\hspace{-0.35ex}\mathbf{f}_{\hspace{-0.2ex}D}(\mathbf{X}_{\hspace{-0.2ex}p}^{(t)}\hspace{-0.35ex})\hspace{-0.3ex}\big|^2\hspace{-0.35ex}\Big). \label{eq:SIC_PSO} \vspace{-1ex}
\end{equation}
At the $t^{th}$ iteration, the individual best for the $p^{th}$ particle as well as the current best among all particles within the swarm are respectively found as follows:
\begin{equation}
	\mathbf{X}_{\mathrm{best},p}^{(t)}= \argmin_{\mathbf{X}_{p}^{(t^*)}, \forall t^* = 0,1,\cdots, t} \text{A}_{\textrm{SI}}(\mathbf{X}_p^{(t^*)}), \label{eq:personal_best_PSOLPA}  
	\vspace{-1ex}
\end{equation}
\begin{equation}
	\mathbf{X}_{\mathrm{best}}^{(t)}= \argmin_{\mathbf{X}_{\text{best},p}^{(t)}, \forall p = 0,1,\cdots, P} \text{A}_{\textrm{SI}}(\mathbf{X}_{\text{best},p}^{(t)}). \label{eq:global_best_PSOLPA} 
	\vspace{-2ex}
\end{equation}
The convergence of the proposed PSO-based joint optimization scheme for enhanced SI suppression depends on the velocity vector $\mathbf{w}_p$ for both personal best $\mathbf{X}_{\mathrm{best},p}$ and global best $\mathbf{X}_{\mathrm{best}}$ solutions, which is defined as:
\begin{equation}
	\mathbf{w}_p^{(t+1)}\hspace{-0.65ex}=\hspace{-0.25ex}\mathbf{\Omega}_1\hspace{-0.15ex}(\hspace{-0.25ex}\mathbf{X}_{\mathrm{best}}^{(t)} \hspace{-0.5ex}-\hspace{-0.5ex} \mathbf{X}_{p}^{(t)})\hspace{-0.35ex}+\hspace{-0.25ex}\mathbf{\Omega}_2(\mathbf{X}_{\mathrm{best},p}^{(t)}\hspace{-0.5ex} -\hspace{-0.5ex} \mathbf{X}_{p}^{(t)})\hspace{-0.15ex}+\hspace{-0.15ex}\mathbf{\Omega}_3^{(t)} \mathbf{w}_p^{(t)}, \label{eq:velocity} \vspace{-1ex}
\end{equation}
where $\mathbf{w}_p^{(t)}$ is the velocity of the $p^{th}$ particle at the $t^{th}$ iteration, $\mathbf{\Omega}_1, \mathbf{\Omega}_2$ are the random diagonal matrices with the uniformly distributed entries over $[0,2]$, and represent the inter-particle relationships and the tendency of each particle to move toward its personal best, respectively. Additionally, we introduce $\mathbf{\Omega}_3=\left(\frac{T-1}{T}\right) \mathbf{I}_{\left(2 N_D + 2 N_U\right)}$, a diagonal inertia weight matrix, which helps find the balance between exploration and exploitation for optimal solution in search space. By using (\ref{eq:velocity}), the position of each particle is updated as:
\begin{equation}
	\mathbf{X}_p^{(t+1)}=\operatorname{clip}\left(\mathbf{X}_p^{(t)} + \mathbf{w}_p^{(t+1)}, \mathbf{X}_{\text{Low}}, \mathbf{X}_{\text{Upp}}\right). \label{eq:position} \vspace{-1ex}
\end{equation}
Here, we have $\mathbf{X}_{\text{Upp}} \in \mathbb{R}^{\left(2 N_D+2 N_U\right)}$ and $\mathbf{X}_{\text{Low}} \in \mathbb{R}^{\left(2 N_D+2 N_U\right)}$, which represent the upper and lower boundaries for the perturbation coefficients, respectively and they are determined based on the predefined limits for each perturbation coefficient specified in $C_1$-$C_5$. To ensure that the coefficients stay within these boundaries, we employ the clipping function, defined as $\operatorname{clip}(y, i, j)=$ $\min (\max (y, i), j)$. Also, unlike sub-optimal approach, we here consider each perturbation coefficient as a continuous variable inside its boundary. The proposed joint optimization scheme for enhanced SI suppression using PSO is summarized in Algorithm 1. 
\begin{figure*}[!t] 
	\centering
	\subfloat[\label{fig:fig2a}]{% 
		\includegraphics[height=5.2cm, width=0.65\columnwidth]{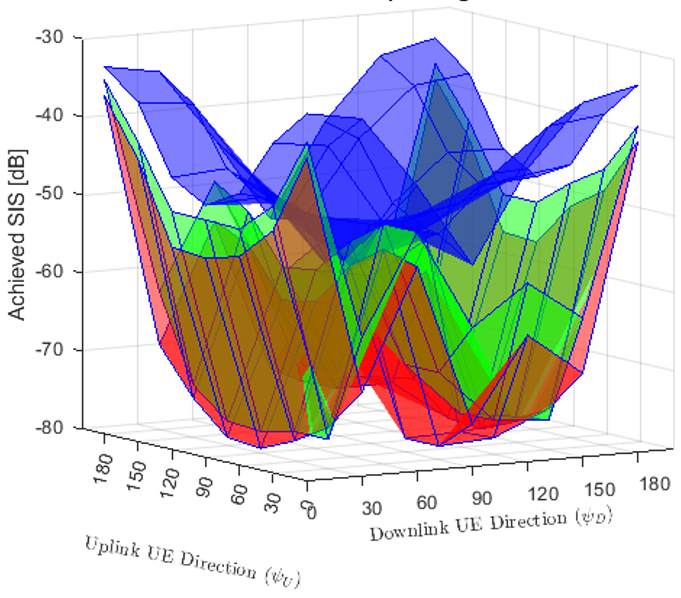} 
	} \hfil 
	\subfloat[\label{fig:fig2b}]{% 
		\includegraphics[height=5.2cm, width=0.65\columnwidth]{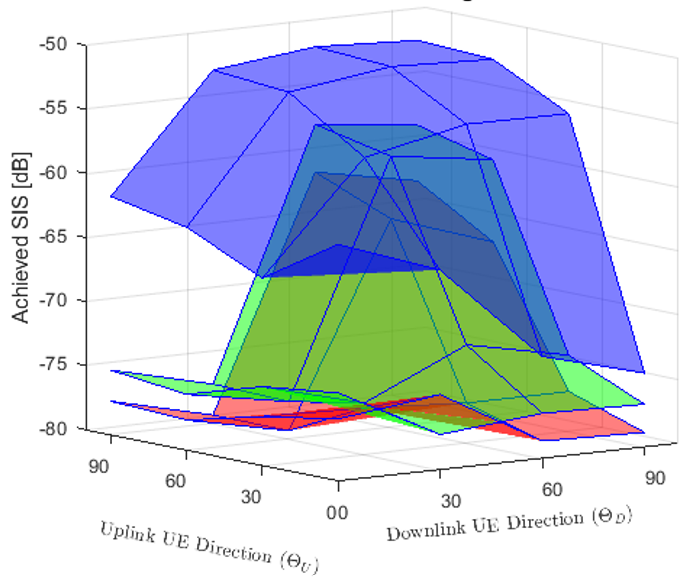} 
	} 
	\subfloat[\label{fig:fig2c}]{% 
		\includegraphics[height=5.2cm, width=0.65\columnwidth]{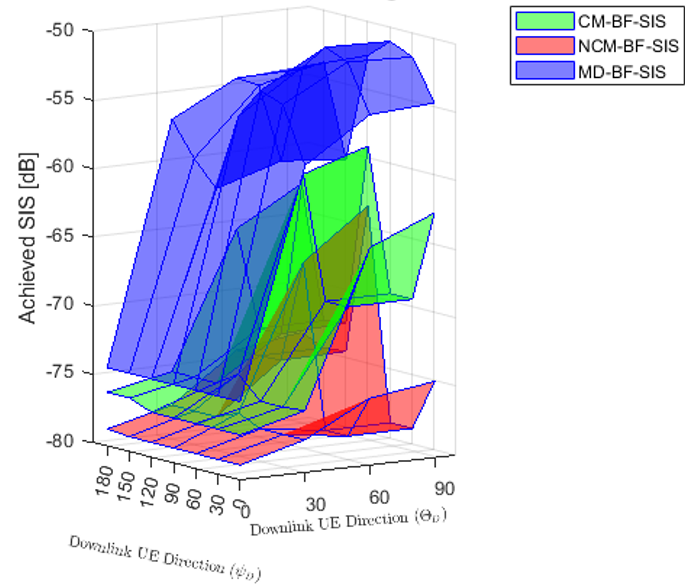} 
	}   
	\caption{Achieved SI suppression of proposed NCM-BF-SIS with 2$\times$2 sub-array at 20 MHz bandwidth. (a) versus $\psi_D$ and $\psi_U$ for fixed ($\theta_D = \theta_U = 90^\circ$). (b) versus $\theta_D$ and $\theta_U$ for fixed ($\psi_D = \psi_U = 90^\circ$). (c) versus $\psi_D$ and $\theta_D$ for fixed ($\theta_U = \psi_U = 90^\circ$).}
	\label{fig:fig2} 
	\vspace{-3ex}
\end{figure*}
\vspace{-1ex}
\section{Illustrative Results}
\vspace{-1ex}
In this section, we present the Monte-Carlo simulation results to illustrate the performance of the proposed SI suppression scheme in FD mMIMO systems. Particularly, we investigate the achieved SI level by the design of RF stages using NCM-BF-SIS scheme. We consider $N_D$ $=$ $N_U$ $=$ 1 RF chain to serve a single uplink and a downlink UE with 2$\times$2 and 4$\times$4 URA sub-array configurations for the results presented hereafter. For PSO, we use $P = 20, \Omega_1 = \Omega_2 = 2$ and $\Omega_3 = 1.1$. In the following, we present the achieved SI suppression results using the proposed NCM-BF-SIS scheme at different uplink and downlink UE angular locations.\par 
%\begin{table*}[t!]
%	\caption{Performance Comparison of Different URA SC-HBF for proposed NCM-BF-SIS scheme at 20 MHz bandwidth. \vspace{-1em}} %title of the table
%	\centering % centering table
%	\resizebox{1\textwidth}{!}{	
%		\begin{tabular}{|c|c||c|c|c||c|c|}
%			\hline
%			\multicolumn{2}{|c||}{\multirow{2}{*}{\shortstack[l]{Sub-Array \\ Configuration}}} & \multicolumn{2}{c|}{Proposed} & \multirow{2}{*}{MD-BF-SIS}  & \multicolumn{2}{|c|}{SI Suppression Gain} \\  
%			\cline{3-4} \cline{6-7} 
%			\multicolumn{2}{|c||}{} & NCM-BF-SIS  & CM-BF-SIS & & NCM-BF-SIS vs MD-BF-SIS & CM-BF-SIS vs. MD-BF-SIS \\ \hline
%			\multirow{3}{*}{$2\times2$} & Max & -79.6 & -77.2  & -74.2 & 16.25 & 33.04 \\ \cline{2-7} 
%			& Min & -38.74 & -40.01 & -61.59 & 0.569 & 1.25  \\ \cline{2-7} 
%			& Avg & -52.6 & -59.69 & -71.96 & 7.07 & 19.33  \\ \hline
%			\multirow{3}{*}{$4\times4$} & Max & 1.930 & 1.929 & 1.948 & 225.8 & 261.1  \\ \cline{2-7} 
%			& Min & 0.514 & 0.597 & 1.598 & -6.18 & -5.887  \\ \cline{2-7} 
%			& Avg & 1.650 & 1.780  & 1.886 & 7.87  & 14.3  \\ \hline
%			\multirow{3}{*}{$1\times4$} & Max & 0.879 & 1.674 & 1.885 & 225.8 & 261.1 \\ \cline{2-7} 
%			& Min & 0.514 & 0.597 & 1.598 & 1.56 & 83.51  \\ \cline{2-7} 
%			& Avg & 0.730 & 1.371 & 1.774 & 87.6 & 143  \\ \hline
%	\end{tabular}}
%	\label{tab:tab0}
%\end{table*}
In Fig. \ref{fig:fig2}, we present the achieved SI using $2\times2$ URA SC-HBF architecture over a bandwidth of 20 MHz, and compare the performance of NCM-BF-SIS with the following two SI suppression schemes: 1) maximum-directivity-based beamforming scheme (MD-BF-SIS), and 2) constant-modulus RF beamforming scheme (CM-BF-SIS)\footnote{In MD-BF-SIS, $\mathbf{f}_D$ and $\mathbf{f}_U$ steer the beams at exact user locations (i.e., $(\theta_D, \psi_D)$ and $(\theta_U, \psi_U$) while in CM-BF-SIS, beam angles are optimized (i.e., $\hat{\theta}_D, \hat{\psi}_D, \hat{\theta}_U, \hat{\psi}_U$) to construct the RF stages with fixed gain controllers.}. In Fig. \ref{fig:fig2}(a), we plot the achieved SI suppression for varying uplink and downlink azimuth locations (i.e., $\{ \psi_D, \psi_U\} \in \{0^{\circ}:30^{\circ}:180^{\circ}\}$) for fixed uplink and downlink elevation angles (i.e., $\theta_D = \theta_U = 90^\circ$). It can be seen that compared to MD-BF-SIS, both NCM-BF-SIS and CM-BF-SIS schemes can provide more uplink-downlink angle-pairs for an enhanced SI suppression. However, the proposed NCM-BF-SIS can achieve the best SI suppression when compared to MD-BF-SIS and CM-BF-SIS schemes, and can bring SI level down to -79.5 dB. The proposed NCM-BF-SIS can further reduce the SI level by around -20 dB and -4 dB versus MD-BF-SIS and CM-BF-SIS schemes, respectively. In Fig. \ref{fig:fig2}(b), we compare the achieved SI for varying uplink and downlink elevation angles (i.e., $\{ \theta_D, \theta_U\} \in \{0^{\circ}:30^{\circ}:90^{\circ}\}$) for fixed $\psi_D = \psi_U = 90^\circ$. The results show that the proposed NCM-BF-SIS scheme can achieve an SI suppression of -75 dB or lower for almost 70 $\%$ of the ($\theta_D,\theta_U$) angle-pairs. Similarly, Fig. \ref{fig:fig2}(c) depicts the SI levels for varying downlink angular locations (i.e., $\theta_D \in \{0^{\circ}:30^{\circ}:90^{\circ}\}, \psi_D \in \{0^{\circ}:30^{\circ}:180^{\circ}\}$) for fixed uplink angular location $\theta_U = \psi_U = 90^\circ$. The proposed NCM-BF-SIS scheme can achieve SI suppression upto -80 dB. Thus, compared to MD-BF-SIS and CM-BF-SIS schemes, jointly optimizing beam angles with gain controller coefficients in NCM-BF-SIS can significantly enhance SI suppression.    \par
\begin{figure*}[!t] 
	\centering
	\subfloat[\label{fig:fig3a}]{% 
		\includegraphics[height=5.2cm, width=0.65\columnwidth]{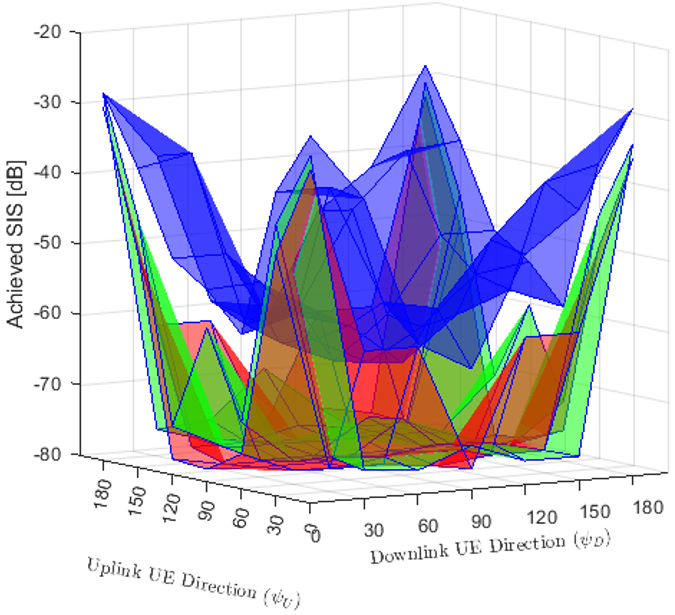} 
	} \hfil 
	\subfloat[\label{fig:fig3b}]{% 
		\includegraphics[height=5.2cm, width=0.65\columnwidth]{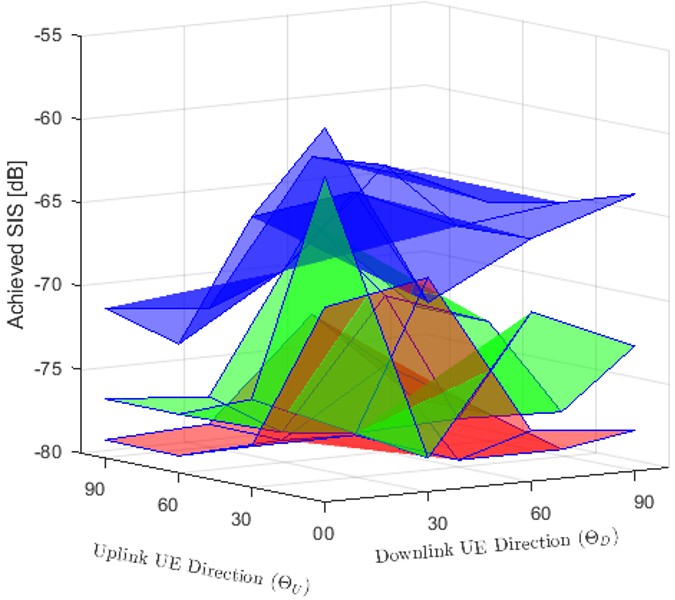} 
	} 
	\subfloat[\label{fig:fig3c}]{% 
		\includegraphics[height=5.2cm, width=0.65\columnwidth]{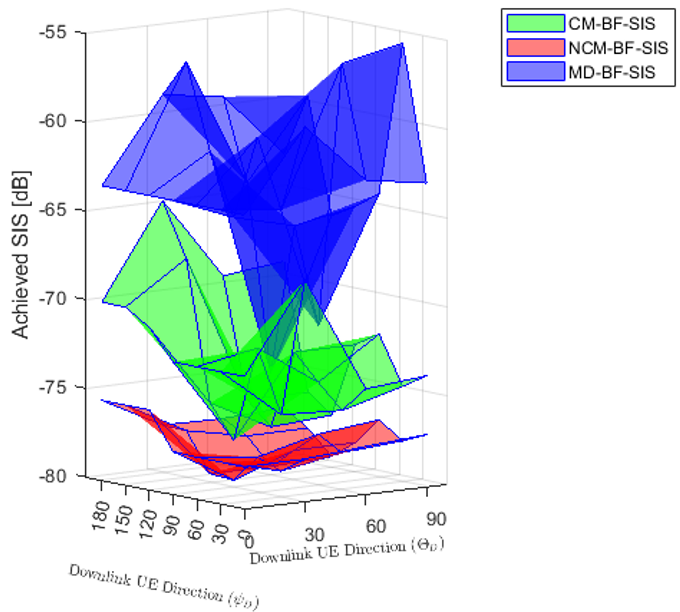} 
	}   
	\caption{Achieved SI suppression of proposed NCM-BF-SIS with 4$\times$4 sub-array at 20 MHz bandwidth. (a) versus $\psi_D$ and $\psi_U$ for fixed ($\theta_D = \theta_U = 90^\circ$). (b) versus $\theta_D$ and $\theta_U$ for fixed ($\psi_D = \psi_U = 90^\circ$). (c) versus $\psi_D$ and $\theta_D$ for fixed ($\theta_U = \psi_U = 90^\circ$).}
	\label{fig:fig3} 
	\vspace{-3ex}
\end{figure*}

 
In Fig. \ref{fig:fig3}, we present the achieved SI using a $4\times4$ URA SC-HBF architecture at 20 MHz bandwidth, and compare the performance of NCM-BF-SIS with MD-BF-SIS and CM-BF-SIS schemes. In Fig. \ref{fig:fig3}(a), the achieved SI suppression for varying uplink and downlink azimuth locations for fixed uplink and downlink elevation angle $\theta_D = \theta_U = 90^\circ$ is presented, which shows the proposed NCM-BF-SIS can achieve the best SI suppression at all $\psi_D-\psi_U$ angle-pairs. Moreover, compared to $2\times2$ URA, a $4\times4$ URA sub-array can provide better performance as the $4\times4$ array can generate narrower beams than $2\times2$ sub-array, which results in enhanced SI suppression. Similarly, Fig. \ref{fig:fig3}(b) compares the achieved SI suppression for different uplink/downlink angle-pairs. The results show that NCM-BF-SIS diminishes the strong SI for any uplink and downlink angle-pair. For instance, the SI can be reduced by -70 dB or lower for any uplink-downlink angle-pair ($\theta_D, \theta_U$) for a fixed $\psi_U = \psi_D = 90^\circ$. In Fig. \ref{fig:fig3}(c), we show the achieved SI levels for different downlink user locations (i.e., $\{ \theta_D, \psi_D\} \in \{0^{\circ}:30^{\circ}:180^{\circ}\}$) for fixed uplink user location $\theta_U = \psi_U = 90^\circ$. It can be seen that the proposed NCM-BF-SIS scheme can attain the SI suppression of -75 dB or lower at all $(\theta_D,\psi_D)$ angle-pairs, whereas MD-BF-SIS scheme can bring the SI level down to -75 dB or lower for only a single $(\theta_D,\psi_D)$ angle-pair. Thus, the proposed NCM-BF-SIS significantly alleviate SI for any uplink-downlink user location (for instance, out of 28 possible ($\theta_D,\psi_D$) angle-pairs, NCM-BF-SIS achieve SI suppression of $\leq$ -75 dB for all 28 pairs, whereas MD-BF-SIS can provide SI suppression of $\leq$ -75 dB for only a single angle-pair). As a result, NCM-BF-SIS can provide more DoF in enhancing FD communications performance irrespective of uplink and downlink user locations. \par 
Finally, Table I presents the detailed comparison of achieved SI suppression using different URA SC-HBF sizes ($2\times2$ and $4\times4$) for NCM-BF-SIS, CM-BF-SIS, and MD-BF-SIS schemes over a bandwidth of 20 MHz. Here, we consider varying uplink and downlink azimuth locations (i.e., $\{ \psi_D, \psi_U\} \in \{0^{\circ}:30^{\circ}:180^{\circ}\}$) for fixed uplink and downlink elevation angles (i.e., $\theta_D = \theta_U = 90^\circ$), and provide the results for best, worst, and average SI suppression. The analysis can be summarized as follows: 1) we can achieve the best SI suppression of around -80 dB by using the proposed NCM-BF-SIS scheme when using either $2\times2$ or $4\times4$ URA sub-array; 2) increasing the number of antennas can provide an enhanced SI suppression (for instance, $4\times4$ URA sub-array can provide more uplink-downlink UE angle-pairs for increased SI suppression than a $2\times2$ URA sub-array); 3) the proposed NCM-BF-SIS can provide an average SI suppression gain of around -20 dB and -4 dB when compared to MD-BF-SIS and CM-BF-SIS schemes, respectively; and 4) the URA sub-array configuration can minimize SI more effectively than ULA sub-array (for example, an average SI suppression of $2\times2$ is around -9 dB lower than $1\times4$ ULA sub-array).
\begin{table}[t!]
	\caption{Performance Comparison of proposed NCM-BF-SIS scheme. \vspace{-1em}} %title of the table
	\centering % centering table
	\resizebox{0.48\textwidth}{!}{	
		\begin{tabular}{|c|c||c|c|c|}
			\hline
			\multicolumn{2}{|c||}{\multirow{2}{*}{\shortstack[l]{Sub-Array \\ Configuration}}} & \multicolumn{2}{c|}{Proposed} & \multirow{2}{*}{MD-BF-SIS}   \\  
			\cline{3-4}  
			\multicolumn{2}{|c||}{} & NCM-BF-SIS  & CM-BF-SIS & \\ \hline
			\multirow{3}{*}{\shortstack[l]{$2\times2$ \\  (URA)}} & Best & -79.6 & -77.2  & -74.2  \\ \cline{2-5} 
			& Worst & -36.6 & -34.6 & -32.2   \\ \cline{2-5} 
			& Avg & -70.2 & -67.3 & -49.3  \\ \hline
			\multirow{3}{*}{\shortstack[l]{$4\times4$ \\  (URA)}} & Best & -79.6 & -78.2 & -74.2  \\ \cline{2-5} 
			& Worst & -36.7 & -35.4 & -33.1   \\ \cline{2-5} 
			& Avg & -72.3 & -69.4  & -54.4  \\ \hline
			\multirow{3}{*}{\shortstack[l]{$1\times4$\\ (ULA)}} & Best & -78.69 & -76.6 & -67.3  \\ \cline{2-5} 
			& Worst & -34.2 & -33.4 & -30.2   \\ \cline{2-5} 
			& Avg & -61.8 & -57.4 & -49.2   \\ \hline
	\end{tabular}}
	\label{tab:tab0}
\end{table}
\vspace{-1ex}
\section{Conclusions}
In this paper, we have considered a FD mMIMO system using a URA SC-HBF architecture for enhanced SI suppression. By utilizing the measured SI channel in an anechoic chamber, we have proposed the non-constant modulus RF beamforming-based SI suppression scheme, which designs the uplink and downlink RF beamforming stages to jointly optimize: 1) \textit{uplink and downlink beam angles}; and 2) \textit{variable gain controllers}, to minimize the strong SI in mMIMO systems. We have proposed a swarm intelligence-based algorithmic solution to find the optimal beam angles combined with gain controllers coefficients while satisfying the directivity degradation constraints in both uplink and downlink directions. The illustrative results show that the proposed NCM-BF-SIS scheme with perturbed beam angles and tuned coefficients can achieve SI suppression as low as -80 dB for FD mMIMO systems. Moreover, compared to ULA sub-array, the URA sub-array can provide an additional SI suppression of around -9 dB.
\vspace{-1ex}
\balance
\bibliographystyle{IEEEtran}
\vspace{-1ex}
\bibliography{references}

\end{document}